\documentclass[preprint2,10.5pt]{aastex}
\begin{document}
\title{\large{\rm{DISCOVERING PROTOSTARS AND THEIR HOST CLUSTERS VIA WISE}}}
\author{D. Majaess}
\affil{{\rm \small Halifax, Nova Scotia, Canada.}}
\email{\rm dmajaess@cygnus.smu.ca}

\begin{abstract}
A hybrid $JHK_s-W_1W_2W_3W_4$ high-spectral index ($\alpha$) selection scheme was employed to identify (sub)clusters of class I/f candidate protostars (YSOs) in WISE observations (the Wide-Field Infrared Survey Explorer). $n>10^4$ candidate YSOs were detected owing to WISE's advantageous all-sky spatial coverage, and a subsample ($n\sim200$) of their heavily-obscured host (sub)clusters were correlated with the \citet{av02} and \citet{di02} catalogs of star-forming regions. Forthcoming observations from the VVV/UKIDSS surveys shall facilitate the detection of additional protostars and bolster efforts to delineate the Galactic plane, since the campaigns aim to secure deep $JHK_s$ photometry for a pertinent fraction of the WISE targets lacking 2MASS detections, and to provide improved data for YSOs near the limits of the 2MASS survey.  
\end{abstract}

\section{{\rm \footnotesize INTRODUCTION}}
Identifying young stellar objects (YSOs) and their host clusters bolsters efforts to constrain the star formation rate, local starburst history \citep[][their Fig.~1]{bb11}, cluster dissolution timescale \citep[`infant mortality rate' for protoclusters,][]{la03}, and the Galaxy's spiral structure.    \citet{bb11} examined newly identified clusters \citep[e.g.,][]{bi03} and inferred that the local star formation rate is not constant and is punctuated for $\tau\leq 9$ and 220-600 Myr, while Spitzer legacy results for five nearby protostar-hosting complexes imply that a sizable fraction of the YSOs lie in loose clusters \citep[$n>35$, $\rho>1 M_{\sun}/{\rm pc}^3$,][]{ev09}.  Such pertinent determinations may be invariably strengthened by increasing the statistics of known protostars and protoclusters.  Hence the importance of infrared surveys such as WISE \citep[][]{wr10}, which facilitate the discovery of such objects \citep{li11,re11,ma12,ko12}. 

Historically, new Galactic clusters were often identified while inspecting photographic plates imaged near optical wavelengths. Young embedded clusters were consequently under-sampled since dust extinction is wavelength-dependent.  By comparison to optical observations, infrared photometry suffers an order of magnitude less dust obscuration \citep[e.g., $A_{[4.5 \mu m]} \sim0.05 A_{V}$,][]{fl07}.   Forthcoming results from the VVV/UKIDSS \textit{near}-infrared surveys \citep{lu08,mi10} are thus pertinent for detecting YSOs and their host clusters, and the observations will extend $\sim 4^{\rm m}$ fainter than 2MASS for Galactic disk stars.  The VVV survey shall establish precise multi-epoch $JHK_s$ photometry for fields in the Galactic bulge and near the Galactic plane \citep[$\ell,|b|\sim294.7,350.0:2.3 \degr$ \& $\ell,b=350.0,10.4:-10.3,5.1 \degr$,][]{mi10,ca11}.  WISE images exhibit a marked improvement in resolution and sensitivity over existing \textit{mid}-infrared surveys (e.g., IRAS), and sample the sky at 3.4 ($W_1$), 4.6 ($W_2$), 12 ($W_3$), and 22 $\mu m$ ($W_4$).  The corresponding FWHM are $6.1 \arcsec$ ($W_1$), $6.4 \arcsec$ ($W_2$), $6.5 \arcsec$ ($W_3$), and $12.0 \arcsec$ ($W_4$).  The Spitzer GLIMPSE surveys \citep[Galactic Legacy Infrared Mid-Plane Survey Extraordinaire,][]{be03,ch09} feature superior resolution relative to WISE, however WISE provides increased (all-sky) coverage.  Extending the GLIMPSE surveys to encompass broader regions of the Galaxy is consequently desirable, and forthcoming.\footnote{\url{http://www.astro.wisc.edu/glimpse/}}

\begin{figure*}
\begin{center}
\includegraphics[width=9.5cm]{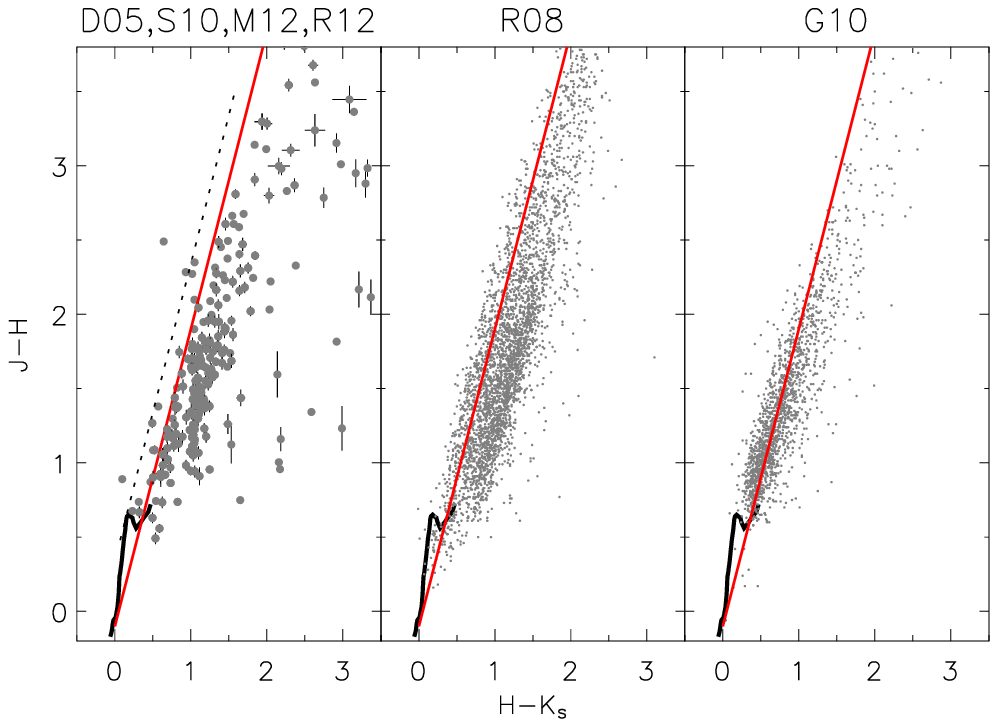} 
\includegraphics[width=6.3cm]{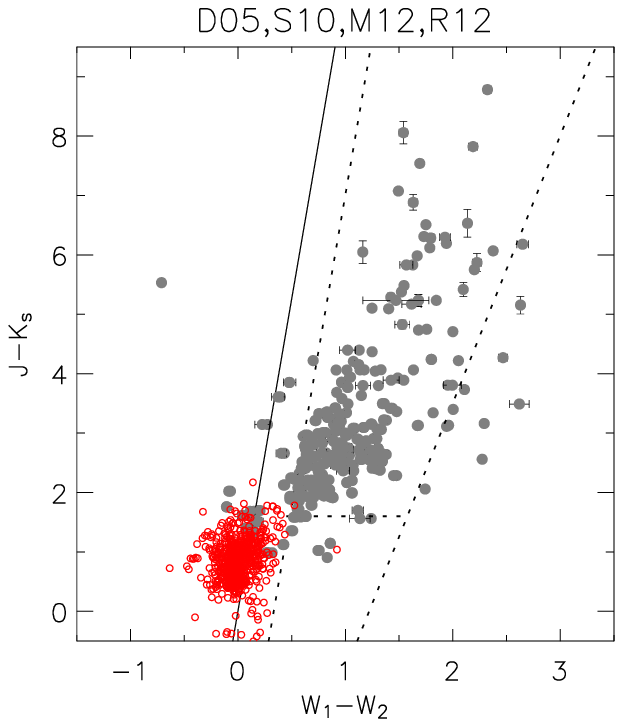} 
\caption{\small{Left, $JHK_s$ color-color diagrams featuring YSO candidates identified by \citet[][D05]{do05}, \citet[][R08]{ro08}, \citet[][G10]{gu10}, \citet[][S10]{sk10}, \citet[][M12]{ma12}, and Rosvick (2012, in prep, R12). The canonical $JHK_s$ reddening law established by \citet{sl08} and \citet{ma11} was adopted.  The black line is the intrinsic relation for main-sequence dwarfs \citep{sl09}, while the dashed line defines the reddening trajectory for red clump stars \citep{sl08,ma11}.  The YSOs lie principally redward of the solid (red) line, and thus that will be adopted as a boundary condition for identifying YSO candidates in the present analysis.  Right, $JK_sW_1W_2$ color-color diagram featuring the YSO samples of D05, S10, M12, and R12.  The YSOs are located primarily within the region bounded by the dashed lines, which will likewise be adopted as boundary conditions for identifying YSO candidates.  The solid line represents the approximate reddening vector for earlier-type stars, while the open red circles define field stars, which typically do not display the signature of IR-excess.  To avoid cluttering the diagrams, errors bars are shown for a subset of the data possessing uncertainties.}}
\label{fig-jhks}
\end{center}
\end{figure*}

\begin{figure*}[!t]
\begin{center}
\includegraphics[width=14cm]{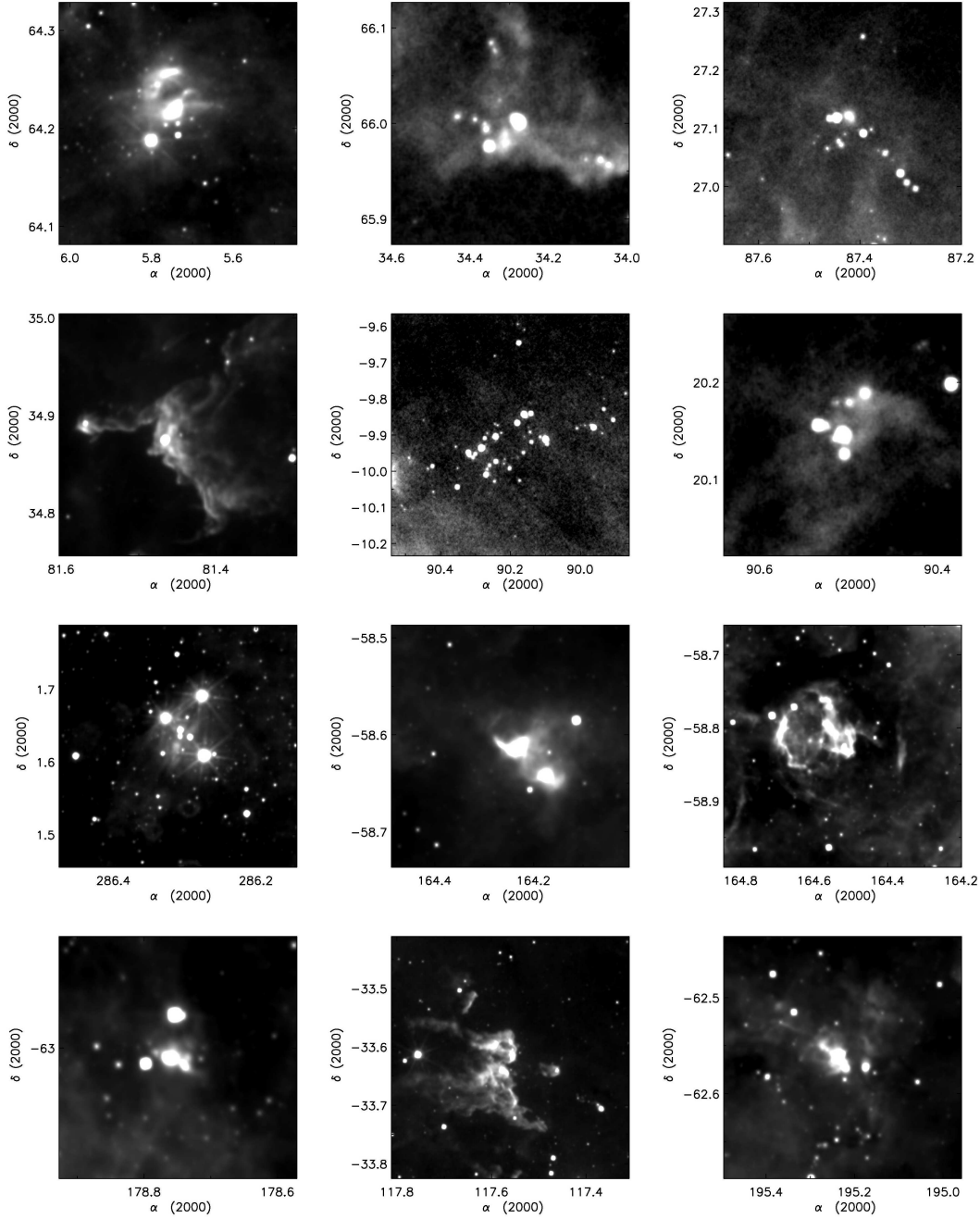} 
\caption{\small{WISE images for a subset of the obscured (sub)clusters detailed in Table~\ref{table1}.}}
\label{fig-cls}
\end{center}
\end{figure*}

The latest generation of infrared surveys are aptly tailored to detect YSOs and their host environments. \citet{ro08}, \citet{ev09}, and \citet{gu10} used Spitzer  data to classify $>13 \times10^3$ YSOs.   \citet{bo11} discovered 96 candidate clusters\footnote{\citet{ch12} discovered numerous Wolf-Rayet stars residing in those clusters using infrared spectra from the VLT, NTT, and SOAR facilities.} in the VVV survey \citep{mi10}, while \citet{me05} identified 92 star clusters via GLIMPSE data \citep[see also][]{fr07,kr06}.  Those infrared surveys resolved numerous individual cluster stars, and in many instances confirmed existing evidence of star formation put forth by low-resolution surveys \citep[e.g., IRAS and maser observations,][]{av02}.  The term discovery is hence somewhat subjective, since a sizable fraction of the aforementioned identifications exhibit entries in the \citet{av02} catalog of star-forming regions, and indeed that is likewise true of the targets described in \S \ref{s-cl}.

In this study, a hybrid $JHK_s-W_1W_2W_3W_4$ high-spectral index ($\alpha$) selection scheme is used to identify YSOs and their host complexes.   This paper is organized as follows: in \S \ref{s-jhkw1w2} 2MASS/WISE color-color cuts inferred from known YSOs, in concert with the slope ($\alpha$) of the spectral energy distribution (SED, \S \ref{s-sed}), is used to identify YSO candidates (\S \ref{s-ysoc}); in \S \ref{s-cl} numerous (sub)clusters hosting the detected YSOs are tabulated, whereby subclusters are offshoot clumps of emerging stars tied to broader star-forming regions (hierarchical clustering); in \S \ref{s-vvv} the pertinence of the VVV/UKIDSS surveys for expanding the YSO sample size is described; and the results are summarized in \S \ref{s-con}. A detailed characterization of individual YSOs \citep[SED modelling,][]{ro07} and protoclusters shall await additional observations (e.g., ALMA), and will be pursued elsewhere.  Ultimately, the results will bolster the statistics linked to supporting new theories of star-formation \citep[e.g., the `fireworks hypothesis',][]{ko12}, and constraining parameters such as the star formation rate and  local starburst history \citep[e.g.,][]{bb11}.

\section{{\rm \footnotesize ANALYSIS}}

\begin{figure*}
\begin{center}
\includegraphics[width=16.5cm]{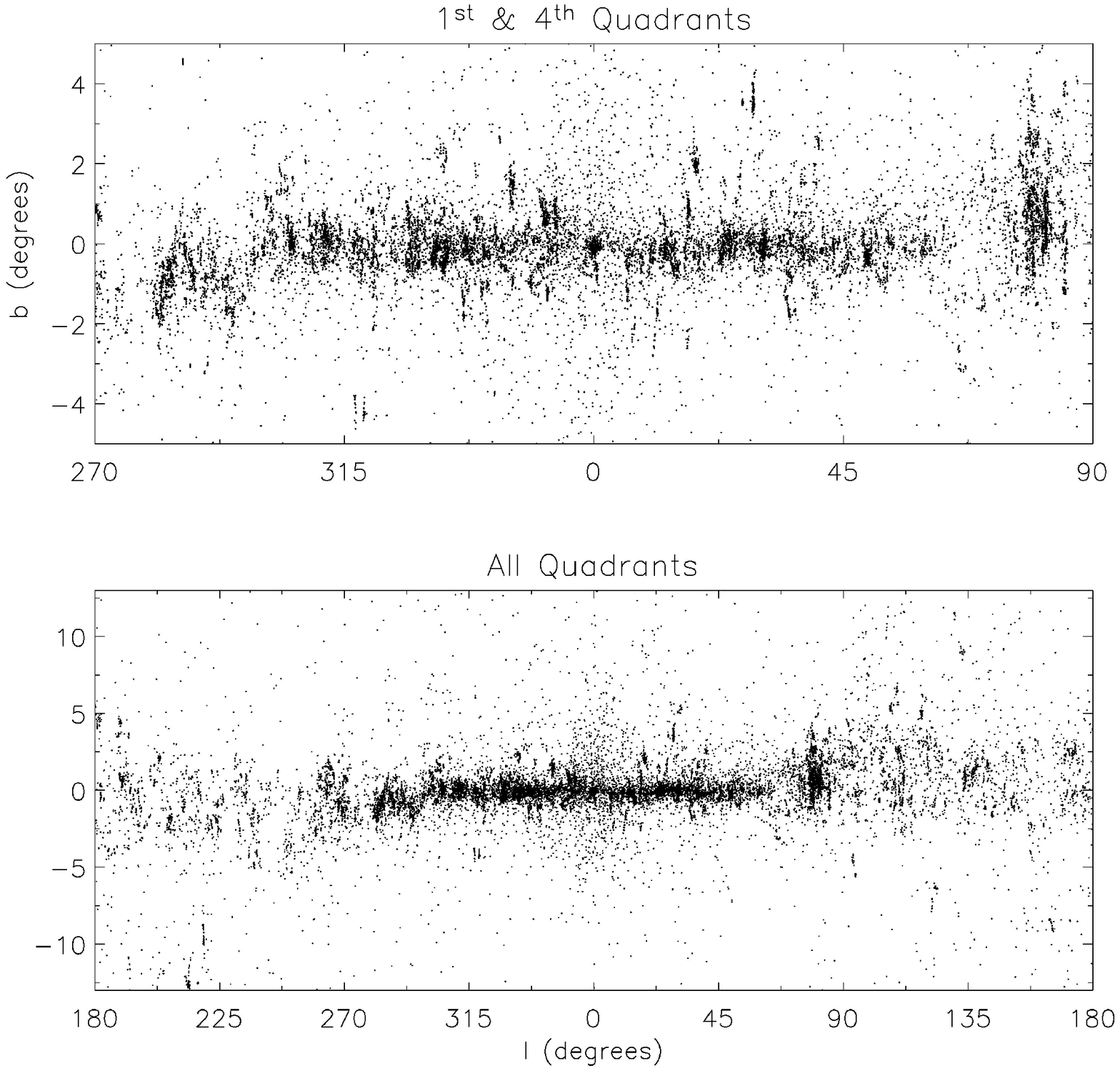} 
\caption{\small{Delineation of the Milky Way via the YSO candidates identified. Large star-forming regions and the warp induced in part by the LMC are discernible.}}
\label{fig-dist}
\end{center}
\end{figure*}

\subsection{{\rm \footnotesize YSO SELECTION SCHEME}}
\label{s-ysoscheme}
\subsubsection{{\rm \footnotesize $JHK_sW_1W_2$ CRITERIA}}
\label{s-jhkw1w2}
 A $JHK_s$ color-color diagram (Fig.~\ref{fig-jhks}) is compiled for the YSO candidates highlighted by \citet{do05}, \citet{ro08}, \citet{gu10}, \citet{sk10}, \citet{ma12}, and Rosvick et al.~(2012, in prep)\footnote{Rosvick et al.~(2012, in prep) detail a new YSO subcluster discovered in $JHK_s$ images acquired from l'Observatoire Mont-M\'{e}gantic \citep[OMM,][]{ar10}. The group was likely triggered by adjacent luminous O-type stars in Berkeley 59 \citep[e.g., the O5V((f))n BD+64$\degr$1673,][see also \citealt{ko12}]{ma08}.  The \citet{do05} results are tied to high-resolution infrared Keck spectra for 41 class I/f YSOs.}. 

Fig.~\ref{fig-jhks} reaffirms that the least evolved YSOs typically occupy positions redward ($H-K_s$) of the reddening line defined by red clump and OB stars. A fraction of the candidates in Fig.~\ref{fig-jhks} lie within the region tied to reddened stars rather than those exhibiting strong infrared excess.  Uncertainties tied to individual passbands add in quadrature and complicate the analysis.  The resulting color-color uncertainties are particularly onerous and can be underestimated for YSOs, which occupy complex encironments and can be detected near the 2MASS survey limits owing to sizable extinction.  The pertinence of the VVV survey for alleviating that problem is discussed in \S \ref{s-vvv}, since the survey extends deeper than 2MASS and exhibits reduced uncertainties for fainter stars.  A photometric cut may be adopted to mitigate field contamination by requiring that relatively unevolved YSOs lie redward of the reddening line for red clump and OB stars, i.e. $(J-H)<E(J-H)/E(H-K_s) \times (H-K_s) - 0.15$ and $(J-H)>1$.   WISE data ($W_1W_2$) may be employed to extend the wavelength baseline and facilitate the detection of infrared excess.  The YSOs identified by \citet{do05}, \citet{sk10}, \citet{ma12}, and Rosvick (2012, in prep) occupy a $JHK_sW_1W_2$ color-color region separated from reddened stars (Fig.~\ref{fig-jhks}).   The following color selection scheme approximately defines that region: $(J-K_s)<10.5 \times (W_1-W_2) - 3.5$, $(J-K_s)>4.5 \times (W_1-W_2) - 5.5$, and $(J-K_s)>1.6$.  Field stars typically do not fall into that regime (Fig.~\ref{fig-jhks}, red open circles). A comparison of low and high-latitude objects passing the aforementioned criteria implies that a magnitude cutoff ($W_3 < 8.7$) reduces contamination by galaxies at larger latitudes.

\subsubsection{{\rm \footnotesize $\alpha$ CRITERION}}
\label{s-sed}
The slope of the SED (spectral index $\alpha$) may be used to facilitate the classification of YSOs.  The canonical framework defines class I, flat, class II, and class III YSOs as featuring $\alpha>0.3$, $-0.3<\alpha<0.3$, $-0.3>\alpha>-1.6$, and $\alpha<-1.6$ respectively \citep[\citealt{gr94}, see also][]{ev09}. However, the wavelength dependence of extinction, in concert with a given YSO's inclination/orientation, introduce degeneracies.   Radiation emitted at 2MASS $K_s$ exhibits increased sensitivity to extinction compared to  WISE passbands
\citep{fl07}.  $\alpha$ is thus sensitive to extinction, and reddened stars may be misclassified as younger protostars \citep[][see also \S 7.2 in \citealt{ev09} and their discussion concerning the Ophiuchus cloud]{ma12}.  Consequently, the spectral index is often employed in tandem with color-color analyses to identify YSOs \citep[e.g.,][]{gu10,ma12}.

Field contamination may be reduced by assessing the positions of high-spectral index candidates in color-color diagrams (Fig.~\ref{fig-jhks}).  High spectral index objects were identified by evaluating the slope of the $
\log{(\lambda  F_{\lambda})}$ function via least-squares (LS) and robust (R) fitting routines.  The fitting routines yield comparable slopes when using 2MASS $K_s+$ WISE photometry, but deviate when relying solely on WISE photometry.  The deduced spectral index is sensitive to the fitting routine and passbands used, and the topic is worth elaborating upon in a separate work (the systematics are not deleterious for the present analysis).   Alternatively, $\alpha$ may be inferred directly from the WISE photometric colors, which is beneficial when 2MASS photometry isn't available: $\alpha_{(LS,W,p)} \sim 0.36(W_1-W_2)+0.58(W_2-W_3)+0.41(W_3-W_4) -2.90$.   Only targets with S/N$>5$ in all WISE passbands were examined, as longer-wavelength 22 $\mu m$ data are valuable for culling non-YSO contaminants \citep[see also][]{ro08}.  $\sim20$\% of the YSO candidates identified toward the Serpens cloud may be reddened giants masquerading as class II/III sources \citep[][and references therein]{ev09}.

To minimize field contamination (e.g., AGB stars) only class I/f objects ($\alpha>-0.3$, \S \ref{s-sed}) are henceforth examined.  Highly reddened field stars (e.g., giants) may exhibit values of $\alpha$ similar to class II/III objects, and indeed, the majority of the AGB stars highlighted by \citet{ro08} peak near $\alpha\sim-0.9$.  Conversely, the YSOs identified by \citet{ro08} peak near $\alpha_{(LS,W,p)}\sim-0.1$.  
\subsection{{\rm \footnotesize YSO CANDIDATES}}
\label{s-ysoc}
High spectral index stars matching the aforementioned $JHK_sW_1W_2W_3W_4$ criteria are classified as YSO candidates.  $\sim10\times10^3$ class I/f YSOs were identified in the VVV survey area, and $30\times10^3$ objects throughout the WISE survey.   The identification of a YSO may be spurious owing to field contamination, photometric uncertainties and blending/crowding (multiple sources falling within the FWHM).  Field contamination appears reduced since $<10$\% of the AGB stars identified by \citet{ro08} were classified as YSO candidates via the present hybrid selection scheme.  \citet{ro08} did not assess YSOs with close neighbors (Fig.~\ref{fig-cls}) in order to mitigate crowding/blending, and adopt $\alpha>-1.2$ as a threshold to detect class II objects.  Only class I/f YSOs were assessed here to reduce field star contamination and an objective was to examine (crowded) protoclusters (Fig.~\ref{fig-cls}).  Admittedly, the criteria adopted here are exceedingly conservative for analyzing obvious YSOs (e.g., class II) in clusters (Fig.~\ref{fig-cls}), and too lax for objects at large Galactic latitudes ($b$) where field contamination (i.e., galaxies) is acute.  Photometric contamination from (non) stellar sources associated with the environment surrounding YSOs will affect the WISE data analyzed, owing in part to the reduced spatial resolution of the observations relative to 2MASS and the matching of the detected sources.  Yet a close-neighbor rejection criterion was avoided in order to achieve the objective of detecting compact groups of YSOs.  Approximately $90$\% of the YSOs identified lack 2MASS neighbors within half the FWHM of the shorter-wavelength WISE passbands.  However, higher-resolution Spitzer photometry via an expansion of the GLIMPSE surveys is desirable.  
  
\subsection{{\rm \footnotesize YSO COMPLEXES}}  
\label{s-cl}
The class I/f YSO candidates identified delineate the Galactic plane as expected (for a comparison to the older PNe distribution see \citealt{ma10}). The ascent from negative $b$ ($\ell\sim270-300 \degr$ to $\ell\sim90 \degr$) is likewise observed in the distribution of classical Cepheids \citep[][see also the \citealt{da01} CO survey]{ma09}.  Distinct conglomerates containing sizable numbers of YSOs are discernible in Fig.~\ref{fig-dist} (e.g., $\ell,b\sim19,2\degr$).  A subsample of the embedded clusters identified, with an emphasis on smaller overlooked subclusters \citep[see also][]{ko12}, are highlighted in Table~\ref{table1} and Fig.~\ref{fig-cls}.  The objects are typically not discernible in optical and even 3.4 $\mu m$ images, which underscores the extreme obscuration.  The bulk of the targets deviate from spherical symmetry and are typically associated with larger complexes (hierarchical clustering).  Constituent stars are observed to emerge from dusty filamentary structure and at the periphery of bubbles \citep[see also][]{ko12}.  The objects were identified while visually inspecting the distribution of YSO candidates (Fig.~\ref{fig-dist})  using the Aladin software environment \citep{bo00}.  Apparent sizes for the (sub)clusters are outlined in Table~\ref{table1}, and those targets tagged by an asterisk contain few members.  The majority of the targets will dissolve prior to achieving open cluster status \citep[][]{la03}.  In many instances the objects are near IRAS and maser sources tabulated in the catalog of star-forming regions \citep{av02}.  The nearest (projected separation) star-forming region lying $r<30\arcmin$ is listed in Table~\ref{table1}, and the offsets between the objects are tabulated.   The (sub)clusters identified were likewise correlated with the \citet{di02} catalog.  The nearest young clusters ($r<20\arcmin$) are listed in Table~\ref{table1}.  The aforementioned catalog is regularly updated, however, the (sub)clusters may be tabulated elsewhere in which case the class I/f YSO members identified here may confirm existing classifications and place solid constraints on the age of the host clusters ($10^5-10^6$ yr).  Clusters which were identified serendipitously as a result of the analysis are likewise tabulated. Table~\ref{table1} shall be made available online in the DAML and WEBDA catalogs \citep{di02,pa08}, as the vast majority of the targets highlighted do not exhibit counterparts in those catalogs.

\begin{figure}[!t]
\begin{center}
\includegraphics[width=6.5cm]{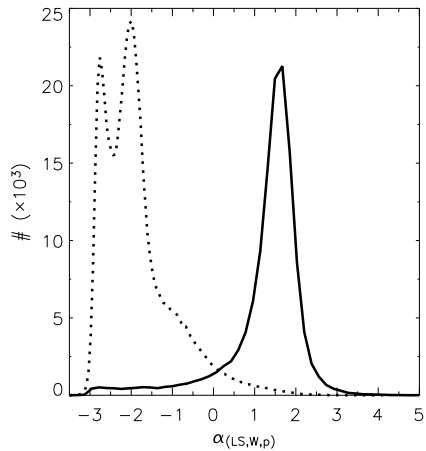} 
\caption{\small{The spectral index ($\alpha_{(LS,W,p)}$) distribution for WISE targets (S/N$>5$) featured in the VVV region (dashed-line).  Stars associated with the maxima exhibit $JHK_s$ colors indicative of late-type giants. The distribution (solid-line) for $>10^5$ stars ($|b|<10\degr$) lacking 2MASS photometry lies principally beyond $\alpha_{(LS,W,p)}>0.3$ (potentially class I YSOs), hence the pertinence of the forthcoming VVV/UKIDSS results (\S \ref{s-vvv}).}}
\label{fig-alphad}
\end{center}
\end{figure}

\subsection{{\rm \footnotesize PERTINENCE OF THE VVV/UKIDSS SURVEYS}}
\label{s-vvv}
A fraction of the class I YSOs identified by \citet{ma12} in the star-forming complex near the classical Cepheid SU Cas lacked 2MASS detections.  Indeed, $>10^5$ objects ($|b|<10\degr$, $W_3<8.7$) featuring S/N$>5$ in all WISE passbands lack 2MASS photometry.  That sample lies principally beyond $\alpha_{(LS,W,p)}>0.3$ (class I, Fig.~\ref{fig-alphad}). One of the main sources of incompleteness for class I YSOs stems from the lack of near-infrared photometry for such objects.    Multi-epoch observations are presently being acquired to complete the full-suite of scheduled VVV ($K_s$) photometry, which may provide photometry for a fraction of WISE targets lacking 2MASS observations, and shall invariably be utilized in concert with longer-wavelength photometry to constrain SED fits \citep[][their Fig.~3]{ro07}.  Work likewise continues on implementing a global PSF (DAOPHOT) photometric pipeline for the VVV survey (Mauro et al. 2012, in prep). VVV images exhibit increased resolution relative to 2MASS, which is important for enabling the discernment of stellar PSFs from material endemic to the (crowded) environments surrounding YSOs (Fig.~\ref{fig-cls}).

\section{{\rm \footnotesize CONCLUSION \& FUTURE RESEARCH}}
\label{s-con}
YSOs and (sub)clusters were identified via a hybrid $JHK_s-W_1W_2W_3W_4$ high-spectral index ($\alpha_{(LS,W,p)}$) selection scheme, namely:  $(J-H)<E(J-H)/E(H-K_s) \times (H-K_s) - 0.15$, $(J-K_s)<10.5 \times (W_1-W_2) - 3.5$, $(J-K_s)>4.5 \times (W_1-W_2) - 5.5$, $(J-K_s)>1.6$, $(J-H)>1$, $\alpha>-0.3$, $W_3<8.7$, and S/N$>5$ in $W_1W_2W_3W_4$ (Fig.~\ref{fig-jhks}). The multiband color-color criteria were inferred from 2MASS/WISE observations for YSOs identified by \citet{do05}, \citet{sk10}, \citet{ma12}, and Rosvick (2012, in prep).  $>30 \times 10^3$ YSO candidates in the preliminary WISE survey were identified.  The objects delineate the Galactic plane and are constituents of giant complexes and highly-embedded (sub)clusters (Table~\ref{table1}, Figs.~\ref{fig-cls}, \ref{fig-dist}).  The impact of field contamination appears mitigated by a selection scheme that requires detections in 7-passbands, as indicated by the identification of protoclusters (Fig.~\ref{fig-cls}, Table~\ref{table1}), the (non-isotropic) confined delineation of the Galactic plane (Fig.~\ref{fig-dist}), and the rejection of the bulk of the AGB sample highlighted by \citet{ro08}.  The present survey is drastically incomplete since it is tied to comparatively shallow 2MASS observations (Fig.~\ref{fig-alphad}).  

The results reaffirm the importance of the latest generation of infrared surveys (e.g., WISE) for enabling the detection of YSOs and their nascent environments \citep[Table~\ref{table1}, Figs.~\ref{fig-cls}, \ref{fig-dist}, see also][]{li11,re11,ko12}.  However, significant work remains and subsequent refinement to the selection scheme (\S \ref{s-ysoscheme}) pending the identification of biases is inevitable, especially given the relative youth of the published WISE data.  Spectroscopic, deep IR photometric, and sub-mm (ALMA) follow-up observations for the cluster targets are desirable, and forthcoming.

\subsection*{{\rm \scriptsize ACKNOWLEDGEMENTS}}
\scriptsize{DM is grateful to the following individuals and consortia whose efforts, encouragement, or advice enabled the research: 2MASS, WISE (D. Leisawitz), C. Bonatto, D. Turner, J. Rosvick, VVV (C. Moni Bidin, D. Minniti, J. Borissova, A-N. Chen\'{e}, D. Geisler, G. Carraro), D. Balam, V. Avedisova, WEBDA (E. Paunzen), DAML (W. Dias), CDS, arXiv, NASA/IPAC ISA, and NASA ADS.  This publication makes use of data products from the Wide-field Infrared Survey Explorer, which is a joint project of the University of California, Los Angeles, and the Jet Propulsion Laboratory/California Institute of Technology, funded by the National Aeronautics and Space Administration.}

\begin{deluxetable}{cllcclc}
\tablewidth{0pt}
\tabletypesize{\scriptsize}
\tablecaption{(sub)clusters}
\tablehead{\colhead{ID} & \colhead{J2000} & \colhead{size ($\arcmin$)} & \colhead{\citet{av02}} & \colhead{offset ($\arcmin$)} & \colhead{\citet{di02}} & \colhead{offset ($\arcmin$)}}
\startdata 
1	&	00:07:21.50 +64:58:22.5			&	5		&	118.29+2.49 	&	$<$1	&		&		\\
2	&	00:09:43.10 +65:20:32.0			&	5		&	118.60+2.81 	&	$<$1	&		&		\\
3	&	00:10:25.80 +65:20:59.6			&	2	*	&	118.60+2.81 	&	4	&		&		\\
4	&	00:10:53.46 +65:27:52.6			&	2	*	&	118.63+3.03 	&	9	&		&		\\
5	&	00:10:57.67 +65:25:12.7			&	3	*	&	118.60+2.81 	&	9	&		&		\\
6	&	00:12:16.69 +60:54:10.1			&	5	*	&	118.62-1.33 	&	28	&		&		\\
7	&	00:14:20.66 +64:29:45.0			&	4		&	118.96+1.89 	&	$<$1	&		&		\\
8	&	00:16:42.51 +64:30:20.6			&	3	*	&	119.20+1.89 	&	$<$1	&		&		\\
9	&	00:21:13.23 63:19:28.5			&	2	*	&	119.56+0.65 	&	$<$1	&		&		\\
10	&	00:22:57.02 +64:12:18.4			&	5		&		&		&		&		\\
11	&	00:23:41.38 +66:13:29.4			&	10		&	120.15+3.38 	&	2	&		&		\\
12	&	00:24:25.30 +65:49:58.2			&	13		&	120.14+3.06 	&	3	&		&		\\
13	&	00:26:16.15 +64:52:30.6			&	7		&	120.36+1.94 	&	$<$1	&		&		\\
14	&	00:28:32.10 65:27:38.0			&	25		&	120.14+3.06 	&	$<$1	&		&		\\
15	&	00:29:21.55 +64:20:02.4			&	3		&	120.54+1.56 	&	$<$1	&		&		\\
16	&	00:29:53.68 +63:51:30.4			&	2		&	120.55+1.20 	&	$<$1	&		&		\\
17	&	00:49:40.53 +65:24:47.4			&	6		&	122.78+2.55 	&	2	&		&		\\
18	&	00:51:26.67 +65:47:42.7			&	7		&	123.20+2.83 	&	$<$1	&	FSR 0516         	&	11	\\
19	&	00:58:28.52 +56:29:53.0			&	14		&	123.13-6.27 	&	3	&		&		\\
20	&	00:58:36.69 +65:40:23.1			&	2	*	&	123.20+2.83 	&	$<$1	&		&		\\
21	&	01:07:54.99 +65:20:25.7			&	18		&	124.64+2.54 	&	$<$1	&		&		\\
22	&	01:08:31.65 +63:08:14.8			&	8		&	124.89+0.33 	&	2	&		&		\\
23	&	01:10:48.27 +63:34:14.0			&	3	*	&	125.09+0.78 	&	$<$1	&		&		\\
24	&	01:15:40.97 +64:46:41.4			&	15		&	125.60+2.10 	&	$<$1	&		&		\\
25	&	01:21:35.52 +62:25:41.4			&	2	*	&	126.66-0.80 	&	23	&		&		\\
26	&	01:45:40.14 +64:16:08.5			&	4	*	&	128.78+2.01 	&	$<$1	&		&		\\
27	&	02:01:18.47 +67:45:33.8			&	4	*	&	129.49+5.77 	&	$<$1	&		&		\\
28	&	02:17:27.02 +65:59:37.1			&	11		&		&		&		&		\\
29	&	02:28:07.00 +72:37:34.7			&	11		&		&		&		&		\\
30	&	02:44:36.37 +60:59:42.4			&	2	*	&	136.09+2.10 	&	$<$1	&		&		\\
31	&	02:54:25.03 +58:10:05.2			&	18		&		&		&		&		\\
32	&	02:58:40.94 +62:26:41.3			&	18		&	137.07+3.00 	&	7	&		&		\\
33	&	03:14:04.91 +58:33:06.9			&	12		&	140.64+0.67 	&	$<$1	&		&		\\
34	&	03:27:31.33 +58:19:21.7			&	5	*	&	142.24+1.42 	&	$<$1	&		&		\\
35	&	03:31:53.53 +60:08:13.3			&	5		&	141.68+3.23 	&	$<$1	&		&		\\
36	&	03:51:36.64 +51:31:00.2			&	4		&	149.09-1.98 	&	$<$1	&		&		\\
37	&	03:53:34.35 +53:36:16.0			&	6		&	148.54-0.24 	&	8	&		&		\\
38	&	03:54:54.65 +53:44:12.9			&	3		&	148.54-0.24 	&	9	&	FSR 0655         	&	14	\\
39	&	03:56:18.21 +53:52:27.0			&	20		&	148.12+0.29 	&	$<$1	&	FSR 0655         	&	$<$1	\\
40	&	03:57:17.44 +54:11:07.6			&	5	*	&	148.04+0.63 	&	$<$1	&	FSR 0654         	&	18	\\
41	&	04:03:19.02 +51:17:57.9			&	25		&	150.58-0.96 	&	$<$1	&		&		\\
42	&	04:03:48.52 +51:01:05.1			&	2	*	&	150.86-1.12 	&	$<$1	&		&		\\
43	&	04:05:53.47 +54:51:04.7			&	7		&	148.50+1.98 	&	3	&		&		\\
44	&	04:07:12.43 +51:23:23.8			&	5	*	&	150.99-0.48 	&	$<$1	&	FSR 0667         	&	13	\\
45	&	04:08:09.67 +50:31:27.4			&	22		&	151.49-1.36 	&	13	&		&		\\
46	&	04:17:54.69 +52:49:40.6			&	10	*	&	151.32+1.99 	&	16	&	Waterloo 1       	&	6	\\
47	&	04:28:28.13 +45:15:01.8			&	18		&	158.48-2.22 	&	23	&		&		\\
48	&	04:36:32.11 +51:13:53.4			&	15		&	154.35+2.61 	&	2	&		&		\\
49	&	04:40:26.46 +60:27:40.5			&	10		&	147.77+9.17 	&	$<$1	&		&		\\
50	&	04:45:29.66 +41:58:33.2			&	17		&	162.28-2.34 	&	$<$1	&	FSR 0721         	&	10	\\
51	&	04:45:45.50 +42:02:05.0			&	16		&	162.28-2.34 	&	5	&	FSR 0717         	&	7	\\
52	&	04:59:11.78 +47:51:19.6			&	7		&	159.16+3.30 	&	9	&	FSR 0696         	&	10	\\
53	&	05:00:23.19 +39:56:33.7			&	14	*	&		&		&		&		\\
54	&	05:16:48.64 +37:01:15.6			&	8	*	&	169.95-0.59 	&	$<$1	&		&		\\
55	&	05:19:01.46 +36:47:33.4			&	7	*	&	170.67-0.27 	&	18	&		&		\\
56	&	05:21:07.27 +36:39:45.1			&	6	*	&	170.67-0.27 	&	$<$1	&		&		\\
57	&	05:21:53.02 +36:38:51.8			&	4	*	&	170.80+0.00 	&	$<$1	&		&		\\
58	&	05:25:51.98 +34:52:30.0			&	14		&		&		&	FSR 0775         	&	6	\\
59	&	05:27:13.61 +38:32:10.8			&	50		&	169.85+1.92 	&	2	&		&		\\
60	&	05:37:23.23 +27:46:20.9			&	13		&	180.03-2.15 	&	$<$1	&		&		\\
61	&	05:38:23.00 +27:26:59.0			&	4		&	180.40-2.13 	&	$<$1	&		&		\\
62	&	05:39:10.00 +27:32:13.2			&	4	*	&	180.40-2.13 	&	12	&		&		\\
63	&	05:40:19.60 +23:52:02.5			&	10		&	183.70-3.64 	&	2	&		&		\\
64	&	05:42:46.26 -09:48:03.9			&	12		&	210.76-19.61	&	12	&		&		\\
65	&	05:49:44.39 +27:06:29.6			&	14		&	182.36+0.18 	&	19	&		&		\\
66	&	05:51:29.90 +27:28:50.0			&	8		&	181.92+0.36 	&	$<$1	&	Dutra Bica 83    	&	9	\\
67	&	05:52:03.29 +27:23:55.6			&	5		&	182.36+0.18 	&	$<$1	&	Dutra Bica 83    	&	$<$1	\\
68	&	05:52:12.90 +26:59:33.0			&	22		&	182.36+0.18 	&	$<$1	&		&		\\
69	&	05:58:13.65 +16:33:33.5			&	15		&	192.16-3.83 	&	2	&		&		\\
70	&	06:00:58.09 -09:54:12.3			&	40		&		&		&		&		\\
71	&	06:02:01.76 +20:08:44.7			&	6	*	&	189.21-1.06 	&	19	&		&		\\
72	&	06:02:08.70 +20:27:47.8			&	9	*	&	189.21-1.06 	&	$<$1	&		&		\\
73	&	06:02:16.88 -09:06:28.8			&	17		&		&		&		&		\\
74	&	06:02:45.97 -09:43:16.8 			&	20		&	216.31-15.05	&	9	&		&		\\
75	&	06:09:44.27 +21:07:03.7			&	3		&	189.68+0.72 	&	12	&		&		\\
76	&	06:10:53.58 +14:09:41.3			&	6	*	&	196.07-3.43 	&	27	&	FSR 0939         	&	13	\\
77	&	06:12:05.34 +20:15:12.6			&	5		&	190.04+0.49 	&	18	&		&		\\
78	&	06:13:35.82 +15:57:39.6			&	4		&	193.69-1.05 	&	22	&		&		\\
79	&	06:27:51.82 +05:31:40.0			&	8		&	206.30-2.11 	&	2	&		&		\\
80	&	06:32:32.03 +10:19:56.0			&	15		&	201.60+0.53 	&	$<$1	&		&		\\
81	&	06:33:15.80 +02:30:22.0 			&	3		&	208.51-3.21 	&	$<$1	&		&		\\
82	&	06:36:39.88 +05:36:01.7 			&	3	*	&	206.26-0.71 	&	$<$1	&		&		\\
83	&	06:58:53.26 -07:45:00.3			&	2	*	&	220.80-1.72 	&	8	&		&		\\
84	&	07:00:34.54 -09:11:52.0			&	4	*	&	221.85-2.02 	&	$<$1	&	Ivanov 4         	&	20	\\
85	&	07:03:26.34 -09:19:56.3			&	40		&	221.85-2.02 	&	19	&		&		\\
86	&	07:18:30.50 -18:22:15.0			&	11		&	231.96-2.06 	&	5	&	ESO 559 02       	&	15	\\
87	&	07:19:35.87 -17:49:10.4			&	50		&	231.96-2.06 	&	$<$1	&		&		\\
88	&	07:24:07.03 -25:53:55.9			&	6		&	237.25-6.50 	&	2	&		&		\\
89	&	07:24:37.71 -24:34:59.4			&	5		&	237.25-6.50 	&	$<$1	&	Ivanov 6         	&	6	\\
90	&	07:24:44.88 -24:29:34.5			&	5		&	237.25-6.50 	&	$<$1	&	Ivanov 6         	&	11	\\
91	&	07:33:15.92 -22:09:19.9			&	14		&	237.26-1.28 	&	$<$1	&		&		\\
92	&	07:34:22.54 -22:37:01.1			&	9	*	&	238.77-1.61 	&	15	&		&		\\
93	&	07:50:14.40 -33:37:07.8			&	25		&	248.97-3.61 	&	10	&		&		\\
94	&	07:51:54.59 -33:14:04.5			&	8		&	248.96-3.21 	&	$<$1	&		&		\\
95	&	08:17:52.55 -35:52:47.6			&	3	*	&	254.05-0.10 	&	$<$1	&		&		\\
96	&	08:18:14.71 -36:03:53.3			&	3	*	&	254.05-0.10 	&	$<$1	&		&		\\
97	&	08:19:10.56 -41:52:04.6			&	3	*	&		&		&		&		\\
98	&	08:20:31.81 -41:51:47.2			&	3	*	&	259.28-2.61 	&	26	&		&		\\
99	&	08:21:44.62 -42:04:55.4			&	10		&	259.61-2.70 	&	17	&		&		\\
100	&	08:22:22.39 -41:36:14.4			&	2	*	&	259.28-2.61 	&	$<$1	&		&		\\
101	&	08:22:48.83 -41:37:09.8			&	2	*	&	259.28-2.61 	&	5	&		&		\\
102	&	08:22:51.06 -41:42:13.7			&	2	*	&	259.28-2.61 	&	8	&		&		\\
103	&	08:23:00.09 -41:55:44.9			&	9		&	259.61-2.70 	&	$<$1	&		&		\\
104	&	08:23:15.31 -41:46:05.7			&	6	*	&	259.61-2.70 	&	10	&		&		\\
105	&	08:24:00.40 -42:24:15.0			&	4	*	&	260.18-3.14 	&	19	&		&		\\
106	&	08:24:41.14 -40:59:57.9			&	7		&	259.29-1.95 	&	16	&	ESO 312 03       	&	18	\\
107	&	08:29:13.96 -41:10:47.7			&	16		&	259.63-1.30 	&	$<$1	&		&		\\
108	&	08:34:20.73 -38:40:28.6			&	6	*	&		&		&		&		\\
109	&	09:03:43.11 -50:28:31.7			&	13		&		&		&		&		\\
110	&	09:07:38.57 -50:41:40.2			&	8	*	&	271.22-1.77 	&	22	&		&		\\
111	&	09:16:10.38 -50:02:59.0			&	4	*	&	271.59-0.53 	&	12	&	Pismis 11        	&	3	\\
112	&	09:18:19.45 -48:26:44.1			&	20		&	270.82+0.69 	&	$<$1	&		&		\\
113	&	09:19:01.86 -46:14:10.1			&	2	*	&		&		&		&		\\
114	&	09:22:20.28 -48:03:58.6			&	4	*	&	271.01+1.39 	&	$<$1	&		&		\\
115	&	09:22:41.44 -48:10:07.1			&	4	*	&	271.01+1.39 	&	2	&		&		\\
116	&	10:03:40.11 -57:26:38.2			&	7		&	281.84-1.59 	&	$<$1	&		&		\\
117	&	10:05:42.66 -57:56:14.9			&	12		&	282.21-2.00 	&	5	&		&		\\
118	&	10:07:30.62 -60:02:38.5			&	2	*	&	283.74-3.41 	&	$<$1	&	Trumpler 12      	&	17	\\
119	&	10:09:27.46 -58:38:51.0			&	10		&	283.55-2.27 	&	24	&		&		\\
120	&	10:10:38.80 -57:45:32.0			&	13		&	282.81-1.34 	&	$<$1	&		&		\\
121	&	10:11:51.12 -58:53:12.6			&	12	*	&	283.55-2.27 	&	7	&		&		\\
122	&	10:12:19.50 -57:34:08.0			&	10	*	&	283.55-0.98 	&	16	&		&		\\
123	&	10:20:56.75 -59:41:06.1			&	40		&	285.04-2.00 	&	11	&	SAI 113          	&	17	\\
124	&	10:26:36.08 -56:33:34.2			&	18	*	&		&		&		&		\\
125	&	10:30:33.41 -58:53:52.0			&	5		&	285.59-0.85 	&	$<$1	&		&		\\
126	&	10:32:36.99 -59:38:48.1			&	10		&	286.40-1.35 	&	11	&		&		\\
127	&	10:33:56.48 -59:43:58.0			&	13		&	286.40-1.35 	&	$<$1	&		&		\\
128	&	10:56:07.77 -60:29:15.9			&	7	*	&	289.07-0.36 	&	10	&	ASCC 63          	&	5	\\
129	&	10:56:26.88 -60:07:42.5			&	22		&	289.07-0.36 	&	$<$1	&	ASCC 63          	&	17	\\
130	&	10:56:59.22 -58:36:44.0			&	9		&		&		&		&		\\
131	&	10:57:41.59 -60:45:45.8			&	6		&	289.41-0.68 	&	10	&		&		\\
132	&	10:58:05.23 -58:49:32.3			&	14		&		&		&	Hogg 9           	&	14	\\
133	&	10:58:42.73 -61:11:14.9			&	4	*	&	289.77-1.30 	&	$<$1	&		&		\\
134	&	10:59:17.09 -60:34:38.9			&	7		&	289.58-0.64 	&	$<$1	&		&		\\
135	&	10:59:35.38 -59:00:01.8			&	4	*	&		&		&	Hogg 9           	&	10	\\
136	&	11:01:00.13 -58:19:18.7			&	2	*	&		&		&		&		\\
137	&	11:01:04.94 -60:51:01.5			&	50		&	289.88-0.75 	&	$<$1	&		&		\\
138	&	11:20:11.34 -62:01:51.8			&	6	*	&	292.92-0.90 	&	29	&		&		\\
139	&	11:24:48.85 -62:13:25.4			&	4		&	293.03-1.03 	&	$<$1	&		&		\\
140	&	11:25:39.64 -62:10:43.8			&	8		&	293.09-0.97 	&	$<$1	&		&		\\
141	&	11:27:29.08 -62:22:55.2			&	10	*	&	293.09-0.97 	&	12	&		&		\\
142	&	11:32:40.59 -62:21:15.7			&	20		&	293.82-0.76 	&	2	&		&		\\
143	&	11:36:44.16 -65:48:45.0			&	10	*	&		&		&		&		\\
144	&	11:54:46.89 -63:07:39.6			&	20		&	296.59-0.97 	&	$<$1	&		&		\\
145	&	11:54:59.78 -62:36:25.5			&	5		&		&		&		&		\\
146	&	11:58:59.03 -63:37:15.9			&	20		&	296.89-1.31 	&	15	&		&		\\
147	&	12:19:55.50 -62:55:04.0			&	8		&	299.30-0.31 	&	$<$1	&		&		\\
148	&	12:19:57.68 -63:45:14.1			&	7	*	&	299.46-1.09 	&	$<$1	&		&		\\
149	&	12:41:10.10 -62:33:48.8			&	3	*	&	302.13+0.29 	&	23	&		&		\\
150	&	12:54:51.86 -61:02:53.7			&	9		&		&		&		&		\\
151	&	12:57:22.68 -61:31:34.2			&	5		&	303.28+1.32 	&	22	&		&		\\
152	&	13:00:54.34 -62:33:46.1			&	20		&		&		&		&		\\
153	&	14:09:51.38 -59:45:58.6			&	5	*	&		&		&		&		\\
154	&	14:12:11.33 -60:56:42.0			&	6	*	&	312.60+0.05 	&	21	&		&		\\
155	&	14:14:13.65 -61:15:44.2			&	7		&	312.60+0.05 	&	7	&		&		\\
156	&	14:14:29.43 -61:12:41.6			&	3	*	&	312.60+0.05 	&	10	&		&		\\
157	&	14:22:03.62 -61:04:04.1			&	3	*	&	313.67-0.12 	&	$<$1	&		&		\\
158	&	14:22:32.41 -61:08:22.6			&	5	*	&	313.67-0.12 	&	$<$1	&		&		\\
159	&	15:00:33.05 -63:13:08.7			&	12		&	314.80-5.20 	&	4	&		&		\\
160	&	15:03:26.41 -63:23:15.1			&	6	*	&	314.80-5.20 	&	24	&		&		\\
161	&	15:19:36.70 -57:19:02.5			&	14		&	321.65-0.03 	&	16	&		&		\\
162	&	15:26:51.68 -56:29:13.5			&	4	*	&	323.46+0.08 	&	15	&		&		\\
163	&	15:29:53.36 -56:35:19.7			&	30		&	323.46-0.08 	&	6	&		&		\\
164	&	15:31:35.89 -56:11:33.5			&	20		&	323.93+0.01 	&	$<$1	&		&		\\
165	&	15:35:16.14 -55:39:31.8			&	13	*	&	324.71+0.34 	&	11	&		&		\\
166	&	15:53:21.08 -55:14:51.6			&	4	*	&	326.86-1.04 	&	5	&		&		\\
167	&	15:58:02.97 -53:57:24.1			&	18		&	328.18-0.59 	&	$<$1	&		&		\\
168	&	15:59:35.20 -52:24:21.6			&	4	*	&	329.46+0.51 	&	$<$1	&		&		\\
169	&	15:59:58.46 -51:37:44.8			&	10	*	&	330.07+1.06 	&	4	&		&		\\
170	&	16:00:51.55 -51:42:40.3			&	5	*	&	330.07+1.06 	&	9	&		&		\\
171	&	16:23:27.20 -49:28:56.8			&	9	*	&	334.17+0.07 	&	$<$1	&		&		\\
172	&	16:29:02.12 -48:59:33.6			&	7	*	&	335.06-0.42 	&	13	&		&		\\
173	&	16:50:50.75 -46:10:43.8			&	5		&	339.72-1.12 	&	$<$1	&		&		\\
174	&	17:00:54.30 -42:19:10.0			&	3	*	&	343.72-0.22 	&	11	&		&		\\
175	&	17:03:25.71 -42:36:05.8			&	7		&	343.93-0.64 	&	$<$1	&		&		\\
176	&	17:04:09.49 -42:28:12.1			&	5	*	&	344.23-0.59 	&	$<$1	&		&		\\
177	&	17:04:14.39 -42:19:57.6			&	12		&	344.23-0.59 	&	$<$1	&		&		\\
178	&	17:11:21.12 -27:25:00.4			&	9		&	357.08+7.19 	&	$<$1	&		&		\\
179	&	17:25:03.87 -37:59:13.0			&	10	*	&	350.01-1.34 	&	5	&	Ruprecht 123     	&	20	\\
180	&	17:28:18.91 -35:04:11.4			&	4	*	&	352.87-0.20 	&	$<$1	&		&		\\
181	&	17:30:18.11 -33:09:18.8			&	19		&	354.66+0.47 	&	2	&		&		\\
182	&	17:31:15.50 -33:52:24.7			&	12		&	354.20-0.05 	&	$<$1	&		&		\\
183	&	17:31:20.10 -33:18:35.4			&	7		&	354.67+0.25 	&	$<$1	&		&		\\
184	&	17:39:17.36 -31:08:46.1			&	5	*	&	357.49-0.04 	&	5	&		&		\\
185	&	17:41:23.84 -30:43:35.9			&	4	*	&	357.99-0.17 	&	$<$1	&		&		\\
186	&	17:54:33.00 -25:52:05.3			&	7		&	  3.66-0.11 	&	$<$1	&		&		\\
187	&	18:06:14.33 -20:31:50.1			&	10		&	  9.62+0.19 	&	$<$1	&		&		\\
188	&	18:07:20.28 -21:52:34.5			&	5	*	&	  8.72-0.51 	&	$<$1	&		&		\\
189	&	18:08:18.27 -20:16:02.6			&	12		&	 10.08-0.09 	&	$<$1	&		&		\\
190	&	18:08:19.78 -22:04:32.1			&	4	*	&	  8.72-0.51 	&	$<$1	&	ASCC 93          	&	11	\\
191	&	18:08:38.58 -19:52:30.5			&	5	*	&	 10.45+0.02 	&	$<$1	&		&		\\
192	&	18:09:01.23 -20:05:06.7			&	18		&	 10.30-0.15 	&	$<$1	&		&		\\
193	&	18:09:09.73 -19:28:38.1			&	20		&	 10.87+0.09 	&	$<$1	&		&		\\
194	&	18:10:28.02 -19:57:09.9			&	16		&	 10.60-0.39 	&	$<$1	&		&		\\
195	&	18:16:52.12 -18:41:00.1			&	10		&	 12.46-1.07 	&	$<$1	&	Turner 4         	&	4	\\
196	&	18:28:23.69 -07:41:01.7			&	8	*	&	 23.45+1.55 	&	$<$1	&		&		\\
197	&	18:59:44.45 +01:01:23.7			&	2	*	&	 35.20-1.75 	&	$<$1	&		&		\\
198	&	19:05:14.43 01:37:17.0			&	9		&		&		&		&		\\
199	&	19:34:45.73 +19:31:52.7			&	3	*	&	 55.16-0.30 	&	$<$1	&		&		\\
200	&	19:34:56.61 +19:14:55.1			&	6	*	&	 55.16-0.30 	&	17	&		&		\\
201	&	19:36:13.21 +20:23:30.4			&	15		&	 56.25-0.17 	&	10	&	FSR 0142         	&	18	\\
202	&	20:05:45.46 +23:25:46.9			&	5	*	&	 62.20-4.53 	&	$<$1	&		&		\\
203	&	20:11:29.49 +40:13:40.6			&	3	*	&	 76.88+3.28 	&	17	&		&		\\
204	&	20:19:48.51 +36:45:50.7			&	6	*	&	 75.22+0.01 	&	20	&		&		\\
205	&	20:20:35.73 +36:50:49.0			&	3	*	&	 75.22+0.01 	&	11	&		&		\\
206	&	20:23:34.49 +36:39:01.1			&	25		&	 75.35-0.43 	&	$<$1	&		&		\\
207	&	20:24:12.98 +35:52:20.6			&	7	*	&	 74.79-0.96 	&	$<$1	&		&		\\
208	&	20:24:39.08 +36:05:45.6			&	5	*	&	 74.79-0.96 	&	14	&		&		\\
209	&	20:37:21.40 +47:14:04.5			&	2	*	&	 85.41+3.74 	&	$<$1	&		&		\\
210	&	20:42:39.24 +48:53:38.3			&	9		&	 87.24+4.05 	&	2	&		&		\\
211	&	21:13:15.44 +46:22:09.0			&	5	*	&	 88.72-1.50 	&	$<$1	&		&		\\
212	&	21:49:40.33 +56:54:40.8			&	1	*	&	100.01+2.36 	&	$<$1	&		&		\\
213	&	22:07:56.62 +59:46:39.0			&	25		&	103.55+3.12 	&	8	&		&		\\
214	&	22:30:00.08 +61:32:55.0			&	4		&	106.90+3.16 	&	$<$1	&	Teutsch 76       	&	10	\\
215	&	22:49:34.51 +59:56:08.6			&	6		&	108.20+0.58 	&	$<$1	&		&		\\
216	&	22:52:42.46 +60:00:04.9			&	4	*	&	108.75+0.25 	&	$<$1	&		&		\\
217	&	22:59:43.08 +62:46:43.6			&	3	*	&	110.15+2.61 	&	$<$1	&	FSR 0413         	&	11	\\
218	&	23:01:22.54 +64:17:21.6			&	2	*	&	111.34+3.92 	&	$<$1	&		&		\\
219	&	23:17:52.60 +58:05:10.0			&	20	*	&	110.78-2.86 	&	17	&		&		\\
220	&	23:18:42.30 +57:44:50.5			&	30		&	110.78-2.86 	&	4	&		&		\\
221	&	23:25:51.86 +64:07:47.0			&	13		&	113.77+2.79 	&	$<$1	&		&		\\
222	&	23:29:07.07 +59:34:19.7			&	7		&	111.73+0.04 	&	27	&		&		\\
223	&	23:30:08.12 +59:25:29.8			&	9	*	&	111.73+0.04 	&	27	&		&		\\
224	&	23:39:17.80 +61:59:14.0			&	6		&	114.61+0.22 	&	4	&		&		\\
225	&	23:39:47.91 +61:55:41.9			&	12		&	114.61+0.22 	&	$<$1	&		&		\\
226	&	23:46:00.23 +59:07:16.8			&	2	*	&	114.61-2.69 	&	$<$1	&	FSR 0443         	&	19	\\
227	&	23:47:18.81 +60:28:03.2			&	13	*	&	115.11-1.44 	&	$<$1	&		&		\\
228	&	23:50:48.85 +63:41:38.8			&	8		&		&		&		&		\\
229	&	23:51:08.19 +63:53:07.8			&	7		&		&		&		&		\\

\enddata
\label{table1}
\end{deluxetable}

\end{document}